# X-ray Study of the Electric Double Layer at the n-Hexane/Nanocolloidal Silica Interface

Aleksey M. Tikhonov*

*University of Chicago, Consortium of Advanced Radiation Sources, and Brookhaven National Laboratory, National Synchrotron Light Source, Beamline X19C, Upton, NY, 11973, USA*

February 24, 2006

*E-mail: tikhonov@bnl.gov

**ABSTRACT**

The spatial structure of the transition region between an insulator and an electrolyte solution was studied with x-ray scattering. The electron density profile across the n-hexane/silica sol interface (solutions with 5-nm, 7-nm, and 12-nm colloidal particles) agrees with the theory of the electrical double layer and shows separation of positive and negative charges. The interface consists of three layers, i.e., a compact layer of $Na^+$, a loose monolayer of nanocolloidal particles as part of a thick diffuse layer, and a low-density layer sandwiched between them. Its structure is described by a model in which the potential gradient at the interface reflects the difference in the potentials of "image forces" between the cationic $Na^+$ and anionic nanoparticles and the specific adsorption of surface charge. The density of water in the large electric field ($\sim 10^9 - 10^{10}$ V/m) of the transition region and the layering of silica in the diffuse layer is discussed.

**INTRODUCTION**

A comprehensive understanding of the structure of the interface of an insulator/electrolyte solution is of fundamental importance in describing electrochemical processes in systems involving membranes, absorbers, catalysts, surfactants, or surfaces of other dielectrics. For example, the interaction of proteins with biological membranes is mediated often by cations of the electrolyte solution (see, for instance Ref. [1]). Due to the solvent's specific interaction with the insulator, a heterogeneous highly polarized region or an electrical double layer forms at the boundary between bulk phases. The properties of the solvent in the transition region can be very different from that in the bulk liquid, which is of considerable interest in electrochemistry, geophysics, and biology.[2-4]

Starting with Gouy and Chapman, suggestions were made about the interfacial structure at the electrolyte/metal interface, and later, at the liquid/liquid interface in terms of the distribution of electrical potential, ionic concentrations, and capacitance.[5-7] These ideas were further developed by analyzing the Poisson-Boltzmann equation under different conditions and system parameters.[8] Many authors theorized about the electrostatics and Gibbs free energy of a charge at the insulator/electrolyte solution interface, mostly using very approximate and rough models[9, 10] (see Refs in [11]). Volkov *et al.*[11] offered a comprehensive insight into the historical development and current status of the double layer theory at the oil/water interface. Experimentally, our knowledge about electrical double layers is mostly based on the macroscopic equilibrium properties of liquid/liquid interfaces, such as interfacial capacitance and surface tension (see, for example Refs [12, 13]). Recently, Luo *et al.*[14] used x-ray reflectivity to study the interface between two electrolyte solutions. They showed that a generalized Poisson-Boltzmann equation along with the potentials of mean force, which are calculated by taking into account the liquid structure, predicts the ion distributions measured in the experiment without any adjustable parameters.

In this article, I report the findings from studies of a transition layer at the interface between n-hexane and colloidal silica solutions, with particle sizes typical for macromolecules (5 – 12 nm) and very large surface-charge densities (0.2 – 0.9 C/m$^2$). Its thickness is comparable with the Debye screening length, which is a typical width of the diffuse layer in the Goy-Chapman theory, between particles of the electrolyte solution (~ 300 – 1000 Å). In accordance with Vorotyntsev *et al.'s*[15] review, the interfacial potential gradient in this system arises due to the difference in the potentials of "image forces" between the cationic Na$^+$ and anionic nanoparticles and the specific adsorption of surface charge.

The n-hexane/silica sol system offers several advantages for x-ray scattering experiments compared with an air/electrolyte- or electrolyte/metal-electrode. First, this oil-water interface has an enhanced structure factor (x-ray reflectivity normalized to the Fresnel function) due to the relatively small difference in the bulk electron-densities of water and n-hexane. Second, scattering from the transition region at the electrolyte/metal is very weak in comparison with Bragg diffraction from the electrode's bulk, whereas scattering from the hexane/silica sol interface is defined by the interfacial structure.[2,3] Finally, the width of the electric double-layer at the hexane/silica sol interface ranges from 15 nm to 40 nm. Consequently, the interfacial structure can be resolved by data with relatively poor spatial resolution compared with those required in the experiments of Toney *et al.*[2] and Wang *et al.*[3]

**EXPERIMENT**

All the data presented in this paper were obtained at the liquid surface scattering spectrometer at beamline X19C, National Synchrotron Light Source, Brookhaven National Laboratory. These experiments used 15 keV x-rays that are adequate for studying interfaces between light oils and water.[16, 17] The planar interface between an immiscible bulk n-hexane and bulk silica solution was studied in a polypropylene sample cell with a circular interfacial area of 100 mm diameter, placed inside a two-stage thermostat. Usually, the x-ray beam illuminated less than 1 % of the area of the interface. The temperature in the second stage of the thermostat was stable to better than $\pm 3 \times 10^{-2}$ K. All x-ray scattering measurements were carried out with samples equilibrated at $T$=298 K for at least twelve hours. At the chosen x-ray wavelength, $\lambda = 0.825 \pm 0.002$ Å, the absorption length for n-hexane is approximately 19 mm.

N-hexane was purchased from Sigma-Aldrich and purified by passing through activated alumina in a chromatography column. DuPont supplied the suspensions of colloidal silica in water. The concentrated sols, stabilized by NaOH, contained silica particles of approximately 50 Å (pH ≈ 10), 70 Å (pH ≈ 10), and 120 Å (pH ≈ 9) in diameter, $D$. The resulting homogeneous solution (30 % $SiO_2$ and 0.5 % Na by weight) had specific gravities, $\xi$, respectively, of 1.1 g/cm$^3$ (16% of $SiO_2$ and 0.3% of Na by weight), 1.22 g/cm$^3$ (30% of $SiO_2$ and 0.5% of Na by weight), and 1.30 g/cm$^3$ (40% of $SiO_2$ and 0.03% of Na by weight). The molar concentration of free hydroxyl ions in the sol bulk is extremely low $c^- \sim 10^{-4} - 10^{-5}$ mol/L compared with the concentration of sodium ions $c^+ = f_{Na} \xi / M_{Na} \approx 2 \times 10^{-1} - 2 \times 10^{-2}$ mol/L ($M_{Na} \approx 23$ g/mol is the atomic weight of Na, and $f_{Na}$ is the weight fraction of sodium in the suspension) due to the adsorption of OH$^-$ ions at the silica surface, which is associated with an energy gain $w^- \sim k_B T \ln(c^+/c^-) \sim 7 k_B T$ per ion ($k_B$ is Boltzmann's constant). Since most of the electrolyte ions are concentrated near silica surface, the Debye screening length in the solution between particles can be as large as $\Lambda_D = \sqrt{\varepsilon_0 \varepsilon_1 k_B T / (c^- N_A e^2)} \approx 300 - 1000$ Å (where $\varepsilon_0$ is the dielectric permittivity of the vacuum, $\varepsilon_1$ is the dielectric permittivity of water in the sol, $e$ is the elementary charge, and $N_A$ is the Avogadro constant).

Alternatively, a particle in the sol can be considered as analogous to a large ion such that the silica sol can be treated as a strong electrolyte in which the solutes are completely ionized. Since $c^- << c^+$, the particles in the sol carry a negative charge $Z \approx e(c^+ N_A/c_b) \sim 400e - 700e$, which corresponds to a charge density, $Q$, at the silica surface of $\sim 0.7 - 0.9$ C/m$^2$ for 5 and 7 nm particles ($Q \sim 0.2$ C/m$^2$ for a 120-Å particle). The bulk concentration, $c_b$, of particles in the suspension was as large as $c_b \sim d_b^{-3} \sim 2\times10^{23}$ m$^{-3}$ for sols with 50-Å and 70-Å particles, respectively, and $\sim 2\times10^{22}$ m$^{-3}$ in the solution of 120-Å particles. The particle-particle distance, $d_b$, was obtained from measuring the small-angle scattering of a bulk sample prepared in 0.5-mm-diameter glass tube.

Since colloidal silica has a gigantic surface-to-volume ratio of $10^7 - 10^8$ m$^2$ per m$^3$, the surface-active impurities present in the sol are mostly adsorbed, as was confirmed by measurements of interfacial tension with a Wilhelmy plate. The tension, $\gamma$, of the hexane/sol interface ranges from 38 to 42 mN/m and is stable within 0.2 mN/m for at least 24 hours, as well as between 10° C to 50° C.

Colloidal suspensions and hexane form a high-contrast interfacial structure. Fig. 1 shows the reflectivity for a system with ~ 70 Å particles. Figs. 2 and 3 depict the structure factor (x-ray reflectivity normalized by Fresnel function) for systems with ~ 50 Å and ~ 120 Å particles, respectively.

Gravity orients the hexane/water interface so that it is useful to describe the kinematics of the scattering in the right-handed rectangular coordinate system where the origin $O$ is in the center of the x-ray footprint; here, the $xy$-plane coincides with the interface between transition region and bulk sol, the axis *x* is perpendicular to the beam's direction, and the axis *z* is directed normal to the interface opposite to the gravitational force (Fig. 4). At the reflectivity condition, $\alpha = \beta$, and $\phi = 0$, where $\alpha$ is the incident angle in the *yz*-plane, $\beta$ is the angle in the vertical plane between the scattering direction and the interface, and $\phi$ is the angle in the *xy*-plane between the incident beam's direction and the direction of the scattering. Since the angles were small in these experiments, the components of the wave-vector transfer *q* at small-angle deviations, $\delta\phi$ and $\delta\beta$, from the specular condition can be written in the following form:

$$q_x \approx \frac{2\pi}{\lambda}\delta\phi;$$

$$q_y \approx \frac{2\pi}{\lambda}\alpha\delta\beta; \quad (1)$$

$$q_z \approx \frac{2\pi}{\lambda}(\alpha+\beta).$$

Reflectivity measurements at small $q_z$ constrain the size and divergence of the x-ray beam incident on the sample.[16] The distance between the center of the sample cell and the closest incident slit is ≈ 120 mm. At the smallest reflection angles, ~ $6\times10^{-4}$ rad ($q_z$ ~ 0.01 Å$^{-1}$), the vertical beam's size must be ~ 15 μm at the sample for the footprint to cover only the interface's flat region (~ 20 mm long). This configuration can be achieved only by reducing natural divergence of the beam, ~ $1\times10^{-4}$ rad, down to ~ $1\times10^{-5}$ rad by the use of two collimating input slits (~ 10 μm gap) separated by ~ 600 mm. At high angles ($q_z > 0.2$ Å$^{-1}$), the maximum vertical size of the input slits, 0.2 mm, is limited by the chosen vertical angular acceptance of the detector, $\Delta\beta = 5.9\times10^{-4}$ rad (0.4 mm slit ~ 680 mm away from the center of the sample). The reflectivity measurements were carried out with the detector's horizontal acceptance $\Delta\phi = 1.4\times 10^{-2}$ rad.

To establish the correct value for the reflectivity of the colloidal systems at small $q_z$ ( < 0.05 Å$^{-1}$), it is very important to carefully account for the parasitic bulk small-angle scattering background (Fig. 3, dots). This value was obtained under slightly off-specular conditions ($q_y = \pm q_z/2000$, $q_x = 0$) and then subtracted from the specular data. The $q$-dependent bulk background also can be observed in the β-scan. The circles in Fig. 5 represent the β-scan taken near $q_z$ = 0.05 Å$^{-1}$, where the strong peak at β = 0.19 deg corresponds to the reflection. There are no peaks in the off-specular scattering associated with the in-plane structure of the interface. Rather, it has the same structure as the small-angle scattering background measured from the bulk sample in the glass tube (dots). Also, in this figure the small-angle scattering background is shifted along the β-axis so that its main peak coincides with the transmission beam at $\beta = -0.19$ deg. Thus, the off-specular peaks can be identified with the bulk scattering peaks. Both sets of the data were taken with the same vertical angular acceptance of the detector, $\Delta\beta = 3\times 10^{-4}$ rad (= 0.017 deg), and input slits gap, 40 μm.

## ELECTRON-DENSITY PROFILE

Several general statements about the reflectivity at an oil/sol interface are useful here. First, x-ray reflectivity contains information about the electron density profile across the interface, averaged over the macroscopically large x-ray footprint on the interface. The structure factor of the air/liquid$_1$ interface can be very different from that of the liquid$_2$/liquid$_1$ interface due to the density difference between the bulk phases (liquid$_1$ is denser than liquid$_2$). Usually, the structure of the adsorbed films at a surface is modeled as a multilayer (see Fig. 6). In standard procedure (see, for example Ref. [18]) the interfacial structure is divided into N layers. Each layer has a thickness $l_j$ and electron density $\rho_j$. In addition, N+1 $\sigma_j$ parameters determine the interfacial width between the layers. The total number of interfaces is N+1. The layers in the N-layer stack are separated by N-1 internal interfaces. The structure factor of the multilayer, in the first Born approximation, is a quadratic form of the electron densities of the layers and bulk phases.[19] An internal interface of the multilayer at a liquid$_2$/liquid$_1$ interface contributes relatively more to the reflectivity structure factor than it does for an air/liquid$_1$ surface by the factor $[\varsigma_1/(\varsigma_1 - \varsigma_2)]^2$, where $\varsigma_1$, $\varsigma_2$ are the bulk electron densities of liquid$_1$ and liquid$_2$, respectively. In particular, for the adsorbed layered structure sandwiched between bulk hexane and water, this factor of contrast enhancement is $\approx 10$, where the bulk electron densities of pure water (liquid$_1$) and hexane (liquid$_2$) are $\varsigma_w = 3.33 \times 10^{29}$ e$^-$/m$^3$, and $\varsigma_2 = 2.26 \times 10^{29}$ e$^-$/m$^3$, respectively. Correspondingly, the contribution of the transition region at n-hexane/water interface to the interfacial structure factor is $(\varsigma_m - \varsigma_w)/(\varsigma_w - \varsigma_2) > 20$ times stronger than it is, for example, at an electrolyte/Ag electrode interface, where $\varsigma_m = 2.76 \times 10^{30}$ e$^-$/m$^3$ is the electron density of bulk silver.

Second, to extract information about the electron density profile from the data, the Parratt formalism was used.[20] Although this formalism gives an exact value for the reflectivity of a given structure, the electron-density profile, established from the reflectivity, is not unique. This ambiguity is connected to the complete loss of phase information for the structure factor and the finite $q_z$ range covered by the measurements. Fortunately, due to the large difference in the densities of hexane and the silica sol, the phase of the structure factor is identical to the so-called Hilbert phase, which is defined by reflectivity only.[21, 22]

Third, the experimental findings for the pure oil/water system showed that the low limit for the parameter $\sigma_0$ was defined by a so-called capillary-wave roughness, $\sigma_{cap}$.[23] Its value is given by the detector resolution, determined by $q_z^{max} = 0.35$ Å$^{-1}$ and a short wavelength cutoff in the spectrum of capillary waves:

$$\sigma_{cap}^2 = \frac{k_B T}{2\pi\gamma} \ln\left(\frac{Q_{max}}{Q_{min}}\right), \qquad (2)$$

where $Q_{max} = 2\pi/a$ ($a \approx 5$ Å is of the order of intermolecular distance), and $Q_{min} = q_z^{max} \Delta\beta/2$. In these experiments the calculated value for $\sigma_{cap}$ is as large as $4.1 \pm 0.2$ Å, which sets the low limit for all $\sigma_j$ parameters. Any additional unspecified intrinsic structure of the interfaces can cause only an increase in $\sigma_j$.

*a) Three-layer model.*

I started with a two-layer model that generated fits of the reflectivity data with a relatively low value of $\chi^2$ (see Table I and the corresponding dashed-dotted lines in Figs. 2, 3). This model fits data for the system with 120 Å particles only at $q_z < 0.1$ Å$^{-1}$ but fails to describe the wide bump of the structure factor at the higher $q_z \sim 0.15$ Å$^{-1}$ (see dashed-dotted line in Fig. 3). This feature is associated with a layer $\sim 10$ Å thick that is also present in the data for the lighter sols. Fitting $\sigma_0$ as an independent parameter resulted in an unreasonably low value ($\sigma_0 \sim 2 - 3$ Å) without improving quality of the fit at high $q_z$. Therefore, I used the resolution limit (2) in its place ($\sigma_0 = \sigma_{cap}$). The two-layer model describes those parts of the structure with the most contrast, that is, the layer with a high density (a monolayer of colloidal particles), and the layer with a low-density of silica.

To get a satisfactory fit at $q_z > 0.1$ Å$^{-1}$ a three-layer model must be employed: the third layer describes the compact layer of cations Na$^+$ (the Stern layer) at the oil/sol boundary. Na$^+$ can be adsorbed due to the effect of the electrical image and to specific adsorption (see Fig. 7). Regardless of the nature of adsorption, the extra ionic density of Na$^+$ can be estimated by adding to the two-layer construction a thin layer broadened by capillary waves.

The following general assumptions can be made about the third layer's structure. First, the minimum value of thickness $l_1^{min} \approx 2$ Å is defined by the diameter of the sodium ion, the smallest particle in the system.[24] Second, the interfacial width between the thin layer ($l_1 \approx \sigma_0$) and the low-density region must be similar to $\sigma_0$ ($\sigma_1 \approx \sigma_0$). Third, the interfacial width of the hexane-water interface is perfectly described by the theory of capillary waves and by the detector's resolution, so that $\sigma_0 = \sigma_{cap}$.[17] Thus, the number of the independent parameters in the three-layer model can be reduced from ten to eight.[25] In Table II $l_1$, $l_2$ and $l_3$ are the thicknesses of the thin layer $\rho_1$, the low-density layer $\rho_2$ and the colloidal monolayer $\rho_3$, respectively. $\sigma_2$ is the interfacial width between the last two, while $\sigma_3$ is the interfacial width between the electrolyte bulk and the colloidal monolayer. The estimated error bars were established either from the uncertainties of the bulk properties, or from the $\chi^2$ distribution versus the number of degrees of freedom, given by the number of data points. Comparing Table I with Table II reveals that the $\chi^2$ parameter for the two-layer model is systematically higher than that of the three-layer model. Figs. 2 and 3 show that the two-layer model fails to describe the reflectivity at high $q_z$ (around 0.15 Å$^{-1}$), a finding consistent with the presence of the thin layer in the structure. The solid lines in Figs. 1, 2, and 3 denote the modeled reflectivity curve and structure factors for the three-layer model. Fig. 8 presents the electron-density profiles.

*b) Resolution-based model.*

There is another approach to fitting the structure factor with the same number of independent parameters as in the three-layer resolution model, which I call a resolution-based analysis that uses a series of models slicing the structure into N layers of the same thickness, *l*. N was varied sequentially from 2 to 8, thereby reducing *l* to the limit of spatial resolution $\sim 2\pi/q_z^{max}$ (for $q_z^{max}$ = 0.35 Å$^{-1}$, $l_{min} \approx$ 20 Å). Assuming that $\sigma_j$ has the same value for all interfaces (except $\sigma_0$ for the oil/water interface), the total number of independent parameters could be reduced to N + 3.

The total thickness of the interfacial structure for both 50- and 70-Å particle suspensions is estimated to be as wide as three diameters of colloidal particles (N$l$ = 200 ± 20 Å, $\sigma$ = 30 ± 3 Å, $\sigma_0$ =4.1 ± 0.2 Å). The solid and dashed-dotted lines in Fig. 9, normalized to the bulk density of water at 298K, depict the Parratt profiles of electron density for the oil/sol interfaces for the resolution-based model. In Fig. 2, the dashed line represents the corresponding structure factor for the 50-Å particle suspension. The quality of the five-layer resolution-based fit for the systems with 50- and 70- Å particles is so good that splitting the structure into more layers does not introduce any new features into the profile.

The thickness of the interfacial structure for the heavier suspension of 120-Å particles is estimated as 3-4 times their diameter (N$l$ = 320 ± 40 Å, $\sigma$ = 30 ± 3 Å, $\sigma_0$ =4.1 ± 0.2 Å). The eight-layer resolution-based model ($l \approx$ 40 Å) fits data only at $q_z$ < 0.1 Å$^{-1}$ and fails to describe the wide bump of the structure factor at the higher $q_z \sim$ 0.15 Å$^{-1}$ (see dashed line in Fig. 3) that is associated with a layer $l_1 \sim$ 10 Å thick. To resolve this feature requires a model with N > 20 containing so many fitting parameters that it is too cumbersome to apply. Therefore, I used the thickness of the first layer, $l_1$, as an independent parameter (N=8) and set $\sigma_1 = \sigma_0 = \sigma_{cap}$. The resulting electron-density profile (dashed line in Fig. 9) is almost identical to that of the three-layer model for this system (see Fig. 8).

Both the three-layer model and more general resolution-based analysis revealed three distinctive layers in the interfacial structure: the layers with a high and low concentration of silica particles, and the thin layer at the boundary between oil and water, which indicates that the negative and positive charges are spatially separated at the interface.

In the first approximation, the colloidal silica is a mixture of water and nanoparticles with bulk content, $f_b$. Thus, the bulk electron-density of silica sol, $\varsigma_1$, is defined by the following composition equation (see Table I):

$$\varsigma_1 = f_b \varsigma_{SiO_2} + (1-f_b)\varsigma_w, \qquad (3)$$

where $\varsigma_{SiO_2}$ is the bulk electron-density of silica particles. Exactly the same equation relates the silica content in the loose monolayer, $f_3$, (or in the low-density layer, $f_2$) with the electron-density $\rho_3$ ($\rho_2$). Assuming a sharp particle-volume distribution (all nanoparticles have almost the same volume), the content of the silica $f = cv$, where $v$ is the volume per particle and $c$ is their volume concentration. Thus, by excluding $\varsigma_{SiO_2}$ from the composition equations, the following very important relationship between the concentration of particles in the monolayer, $c_3$, (or in the low-density layer, $c_2$) and $\rho_3$ ($\rho_2$) is obtained:

$$\frac{c_{2,3}}{c_b} \approx \frac{\rho_{2,3} - \varsigma_w}{\varsigma_1 - \varsigma_w}, \qquad (4)$$

Values of $c_{2,3}/c_b$ for the resolution-based model calculated by Eq. (4) are ~ 30 % lower than those in the three-layer model (listed in Table III).[26]

Since the concentration of silica in the low-density layer is small, it is reasonable to assume, in the first approximation, that the electron-density of the thin (compact) layer reflects the mixture of $Na^+$ and water only (hydrated sodium ions). Both the water molecule and sodium ion have ten electrons. Then, the surface concentrations of $Na^+$, $\Gamma^+$, and water, $\Gamma^w$, in the thin layer can be estimated from the following constraints on the number of electrons in the layer and its volume per unit area:

$$\begin{cases} \Gamma^+ + \Gamma^w \approx 0.1\Gamma; \\ \Gamma^+ V^+ + \Gamma^w V^w \approx l_1, \end{cases} \quad (5)$$

where $\Gamma$ is the integral number of electrons per unit area in the first layer ($\Gamma \approx l_1 \rho_1$), $V^+ \approx 4$ Å$^3$ is the volume of $Na^+$.[24] $V^w$ is the volume per H$_2$O molecule in the layer. In the next approximation $V^w$ can be treated as the volume per 10 electrons of the solvent (for example, mixture of water and silica) with the average electron-density $\rho_2$. $V^w$ is 3% less than the volume per H$_2$O molecule in the bulk water $V_0^w$ ($V_0^w \approx 30$ Å$^3$) for a 50-Å sol, and up to 15 % for a 120-Å sol ($dV/V = -d\rho/\rho$).

When both $\delta V^w / V_0^w = (V^w - V_0^w)/V_0^w$ and $V^+/V_0^w$ are small, the following equation can be obtained from (5):

$$\Gamma^+ \approx \left[ 0.1\Gamma - \frac{l_1}{V_0^w} \right] \left[ 1 + \frac{V^+}{V_0^w} \right] + \frac{l_1}{\left(V_0^w\right)^2} \delta V^w. \quad (6)$$

Values of $\Gamma^+$ for the resolution-based model calculated by Eq. (6) are two to three times higher than those in the three-layer model (listed in Table III).

**ELECTRICAL DOUBLE LAYER**

It is very interesting to relate the observed structure to the properties of the electrical double layer, the theory of which predicts different planes of closest approach to the interface for different components of the electrolyte solution. Using Tables II and III, some physical characteristics of the electrical double layer can be evaluated. The low-density region, as wide as two diameters of the colloidal particles sandwiched between the compact and the diffuse layers, defines Helmholtz's capacitance of the structure as $\sim \varepsilon_0 \varepsilon / d \sim 0.01 - 0.1$ F/m$^2$. Here, an effective dielectric permittivity of the adsorbed layer is assumed to be $\varepsilon \sim 10 - 80$ and the layer's thickness is $d \sim 100$ Å. The values were close to the deferential capacitance measured by impedance techniques for various systems with an electrical double layer.[11] Since the sodium layer is situated at a distance $\sim \sigma_0$ from the oil/water interface, Stern's correction to interfacial capacitance can be ignored, and the interfacial potential drop, $\psi$, can be established immediately from the Goy-Chapman theory $\psi \approx 2k_B T / Z \sim 10^{-4}$ V. Applying this theory further to describe the thick adsorbed layer is problematic, and a more sophisticated model is required.

In current electrochemistry, the formation of the electrical double layer at the dielectric/electrolyte solution usually is explained by the following factors: a spontaneous polarization of the media near the boundary; positive and negative adsorption due to the effect of the electrical-image forces; the specific adsorption of surface charge; and, a nonzero space charge in the adsorbed layer. Vorotyntsev *et al.*[15] reviewed the general problem of distribution of the interfacial potential drop for a thick transition layer wherein a fixed space charge, the solvent molecules, and ions are in equilibrium with the electrolyte. They suggest visualizing the surface of hexane and the Helmholtz plane for nanoparticles as two individual interfaces, contributing independently to the drop in potential across the interface.

First, according to previous studies of a pure system, the polarization of the media at hexane/water interface is not strong enough to create any ordered structure. The interfacial structure can be described only by a spectrum of capillary waves.[17]

Second, the effect of ``image forces'' arises from the different dielectric bulk properties of the phases in contact, and from the inhomogeneous transition region between hexane and the silica sol. The transition region is due to the different planes of closest approach to the interface for the colloidal particles and Na$^+$ that can be understood from the "classical" single-particle energy of interaction with the electrical image[27]

$$\frac{Z^2}{16\pi\varepsilon_0\varepsilon_1}\frac{\varepsilon_1-\varepsilon_2}{\varepsilon_1+\varepsilon_2}\frac{1}{h}, \qquad (7)$$

where $Z$ is a charge of the particle (ion), $\varepsilon_1 = 78$, and $\varepsilon_2 = 2$ are the dielectric permittivities of water and hexane, respectively, and $h$ is the distance from the center of the particle (ion) to the interface. Eq. (7) does not account for the polarization of the interface and the changes of the dielectric properties of the media in the transition region although it explains quantitatively the main effect. The large $Z$ of the silica particle keeps it far from the interface to minimize energy (7). On the other hand, the plane of the closest approach for the sodium ions or outer Helmholtz's plane can reside very close to the oil boundary, so that the thickness of the ion-free layer is about the size of a water molecule.[28] Thus, the charge separation at the interface is unavoidable, but, for the following reasons, the structure cannot be explained only by the effect of "image forces".

Third, assuming that all the negative countercharge is concentrated in the monolayer of silica particles, the condition of the electrical neutrality gives $c_3 \approx e\Gamma^+/(DZ)$, where $Z \approx e(c^+ N_A/c_b)$. Thus,

$$\frac{c_3}{c_b} = \frac{L}{D}, \qquad (8)$$

that is only slightly higher for the system with 50-Å and 70-Å particles but it is three to five times more than that determined from the electron density profile for the solution of 12 nm-particles (see Eq. (4) and Table III). Colloidal particles in this layer must carry a higher charge than those in the bulk to stabilize the in-plane structure, aided by the additional adsorption of hydroxyl ions into this layer.

Finally, the space charge in layer 1 and layer 2 is due to the spatial distribution of $Na^+$. A simple estimation using the bulk properties of the 50 Å- and 70 Å-particle sols shows that a layer as wide as $L = \Gamma^+/(c^+ N_A) \sim 300$ Å near the interface must be deficient in sodium to create the compact layer ($L \sim 600$ Å for the resolution-based model). $L$ is wider than the thickness of the interfacial structure (~ 200 Å). For a suspension with 120 Å particles, $L \sim 2000$ Å is even wider due to a lower $f_{NA}$. Sodium can infiltrate into the compact layer (Stern layer) due to its specific adsorption (reversible ionization of the hexane surface) caused by non-Coulombic short-range forces, and thereby form a compact or loose monolayer. There, the space charge density is low, and the Gouy-Chapman theory can describe the potential distribution within layers 1 and 2 near the boundary with oil.[15]

This seeming failure of the electro-neutrality of the system with 12-nm particles demonstrates the distinction of the interface from the bulk sol. The redistribution of the charge at the interface is not just limited to the spatial rearrangement of the particles and ions, but also involves significant charge transfer from the bulk of the solution that serves as a reservoir of $Na^+$ and $OH^-$.[15] The supporting evidence for specific adsorption is given in Table II which shows that the fitted values for the $\Gamma^+$ do not fully depend on the sol in contact with hexane, while the concentration of sodium in the sols differs tenfold.

A model without specific adsorption of sodium at the hexane surface cannot explain the profile of electron density. In this case, the space charge of layers 1 and 2 would be associated mainly with the ionic concentration of $Na^+$. Therefore, the electric field would be zero at the oil boundary but at a maximum in layer 2, so that its electron density due to electrostriction would be greater than it is for layer 1, thereby contradicting the experimental findings.[4]

The structure of layer 1 and 2 might be more complex if the charge in the compact layer was screened by a diffuse layer of $OH^-$ located near the surface with hexane. However, this situation is unlikely: there is a huge deficit of free hydroxyl ions in the solution that would prevent the buildup of any significant countercharge in layer 1. Unfortunately, before such a model that describes, for example, a variation of electrolyte concentration in the "surface water" layer could be tested, the spatial resolution of the x-ray scattering experiment would have to be improved significantly.

**DISCUSSION**

*a) Compact layer*

The strongest interaction in this system is associated with the repulsion of the nanoparticles from each other and from the oil surface by the forces of electrical image that are $\sim Z^2$. The energy gain, $w$, of the adsorption of nanoparticles at the monolayer is very small: $w \sim k_B T \ln(c_3/c_b) \sim 0.2 k_B T - 0.6 k_B T$ per particle. On the other hand, the interaction of the compact layer is attractive, $\sim Z$, due to either the nanoparticles or the "image charge" induced by them. The adsorption of $Na^+$ in the compact layer is associated with the energy gain, $w^+$, which is comparable to $w^-$: $w^+ \sim k_B T \ln(c_1^+/c^+) \sim 5 k_B T - 7 k_B T$ per ion ($c_1^+ = \Gamma^+/l_1$ is the volume concentration of sodium in the compact layer). The mutual repulsion of the cations in this layer is $(Z/e)(a/d) \sim 10 - 30$ times smaller than their attraction to the nanoparticles ($a \sim 1/\sqrt{\Gamma^+} \sim 5$ Å is the average distance between ions in the layer). Therefore, the compact layer of sodium ions at oil/silica sol interface can be treated as a two-dimensional system similar to an electron gas at a semiconductor surface that is in contact with an insulator (e.g., $Al_xGa_{1-x}As$ heterostructures).[29] According to Vorotinsev and Ivanov, adsorbed ions at high densities can be in a solid state with an area per ion $\sim 10 - 20$ Å$^2$.[30]

*b) "Surface water"*

According to Danielovich-Ferchmin and Ferchmin, in a very strong electric field of $E > 10^8$ V/m, the density of water is significantly higher than it is at normal conditions due to the ordering of dipole moments of $H_2O$ along the field, $E$. A decade ago, several authors obtained disparate results from measuring the density of water near the surface of a metal electrode. In an x-ray reflectivity study of the Ag electrode surface in contact with NaF electrolyte solution Toney *et al.*,[2] reported that the density of the first two to three layers of water near the electrode surface was very high. However, according to Wang *et al*.,[3] the density of "surface water" at an Au electrode did not differ much from the density of bulk water.

Specific adsorption depletes the entire transition layer of sodium ions, so considerably increasing the Debye screening length in layer 2. The electric field, $E$, which, in the first approximation, can be considered as a constant $E = \Gamma^+/\varepsilon\varepsilon_0 \sim 10^9 - 10^{10}$ V/m ($\varepsilon$ is the dielectric permittivity of water in the layer, $\varepsilon < \varepsilon_1$), may significantly change the water density in layers 1 and 2 by electrostriction.[4] In fact, the values of $\rho_2$ in Table II deviate from $\varsigma_w$ by less then 15 %. This result agrees with work of Wang *et al.*[3] An earlier report explored the density of water in a solution of 70 Å particles using x-ray reflectivity and small-angle grazing incidence diffraction.[26]

*c) Layering of silica in a diffuse layer*

Earlier, Madsen et al.[31] used x-ray scattering to study the interface of an air/silica solution of unspecified alkalinity containing particles larger than ~ 300 Å in diameter. Their model for the surface-normal structure in the electron density profile, based on data with relatively poor spatial resolution, $2\pi/q_z^{max} > 100$ Å, postulated three layers of silica particles near to the surface. This type of layering cannot explain the data presented here, either the reflectivity data at high $q_z$, or the angular dependence of the grazing incidence small-angle scattering at the n-hexane/silica sol interface.[26]

Although the dielectric permittivity of the solution is very inhomogeneous across the transition region, in fact there are layers of the solvent where it is constant (see Fig. 7). Therefore, some layering of the nanoparticles below the Helmholtz plane seems possible. For example, particles in the bulk will be repelled from the boundary between low-density layer and loose monolayer by "image forces". This effect could explain the profile of the resolution-based analysis that shows lower density on the both sides of the loose monolayer (see Fig. 9).

*d) Width of the transition region*

The forces between sol particles, cationic $Na^+$, and the charge density induced by them near the interface define the equilibrium structure of the interface. The transition layer thickness of the 50-Å and 70-Å-particle solutions is the same, while it is much wider for the 120-Å sol (200 Å vs. 400 Å). As shown above, 12-nm particles in the loose monolayer should carry at least a threefold higher charge than in the bulk of the solution to satisfy the condition of electro-neutrality at the interface. This means that the repulsion between particles in the bulk and those at the interface is weaker than the repulsion between the image charge and particles at the interface. On the other hand, the interaction of nanoparticles in the monolayer with the image charge should also be decrease with increasing distance from the interface due to screening by electrolyte ions ($\Lambda_D \sim 1000$ Å). Therefore, the Helmholtz plane for 12-nm particles must be positioned further from the interface than for 5-nm or 7-nm particles.

ACKNOWLEDGEMENTS

Brookhaven National Laboratory is supported by U.S.D.O.E., Contract No. DE-AC02-98CH10886. X19C is partially supported through funding from the ChemMatCARS National Synchrotron Resource, the University of Chicago, and the University of Illinois at Chicago. I thank Professors Mark L. Schlossman and Vladimir I. Marchenko for valuable discussions. I also thank Avril Woodhead.

Å) for the particle-particle distance near the hexane/suspension interface. More importantly, the angular dependence of the small-angle scattering can be explained by the inhomogeneous interfacial structure that contains the plane of the closest approach to the interface for the nanoparticles, above which the concentration of particles is at least ten times lower than that in the bulk ($c_2/c_b < 0.1$). This is in very good agreement with the interfacial models discussed in this paper [A. M. Tikhonov, *J. Phys. Chem. B*, **110**, 2746 (2006)].

Figure 1. X-ray reflectivity for the n-hexane/silica sol interface. The colloidal particles in the suspension are ~ 70 Å. The solid line represents the three-layer model.

Figure 2. Structure factor of the n-hexane/silica sol interface. The colloidal particles in the suspension are ~ 50 Å. The critical angle is $\alpha_c \approx 0.05 \deg$ ($q_c \approx 1.3 \times 10^{-2}$ Å$^{-1}$). The solid line is the three-layer model; the dashed line is the resolution-based model; the dashed-dotted line is the two-layer model.

Figure 3. X-ray reflectivity (open circles) and off-specular background ($q_y = -q_z/2000$, $q_x = 0$) (dots) for the n-hexane/silica sol interface normalized by Fresnel function. The colloidal particles in the suspension are ~ 120 Å. The critical angle is $\alpha_c \approx 0.06 \deg$ ($q_c \approx 1.6 \times 10^{-2}$ Å$^{-1}$). The solid line is the three-layer model; the dashed line is the resolution-based model; the dashed-dotted line is the two-layer model.

Figure 4. Kinematics of the scattering at the hexane/water interface. The *xy* plane coincides with the interface, the axis *x* is perpendicular to the beam's direction, and the axis *z* is directed normal to the interface opposite to the gravitational force. **k**$_{in}$ and **k**$_{sc}$ are, respectively, wave-vectors of the incident beam and beam scattered towards the point of observation, and **q** is the wave-vector transfer, **q** = **k**$_{in}$ - **k**$_{sc}$.

Figure 5. The small-angle scattering (dots) and off-specular $\beta$-scan (circles) backgrounds for the ~ 120 Å particle suspension. The distance between the main peak and the principal ring in the small-angle scattering is $0.15 \pm 0.01$ deg, corresponding to the particle-particle distance $d_b$ ~ 400

Å in the sol. The scan is shifted along the $\beta$-axis so that the transmission beam (the main peak) is at $\beta = -0.19$ deg (not shown). The strong peak at $\beta = 0.19$ deg in the off-specular scan corresponds to the reflection at $q_z = 0.05$ Å$^{-1}$. The $\beta$-scan was measured with the detector's vertical angular acceptance at $\Delta\beta = 3\times10^{-4}$ rad (= 0.017 deg), and the horizontal acceptance $\Delta\phi = 1.4\times10^{-2}$ rad (= 0.8 deg). The lines were drawn by eye.

Figure 6. The interfacial structure is divided into N layers. Each layer has a thickness $l_j$ and electron density $\rho_j$. In addition, N+1 $\sigma_j$ parameters determine the interfacial width between layers.

Figure 7. Three-layer model of the transition layer at n-hexane/silica sol interface.

Figure 8. The profiles of electron density normalized to $\varsigma_w = 3.33\times10^{29}$ e$^-$/m$^3$ across the n-hexane/silica sol interface based on the model of the three-layer structure for the ~ 50 Å - (dashed-dotted line), ~ 70 Å-suspensions (solid line), and ~ 120 Å-particle suspensions (dashed line).

Figure 9. The profiles of electron density normalized to $\varsigma_w = 3.33\times10^{29}$ e$^-$/m$^3$ across the n-hexane/silica sol interface for the resolution-based model for particle suspensions of ~ 50 Å (dashed-dotted line, N=5), ~ 70 Å (solid line, N=5) and ~ 120 Å (dashed line, N=8).

**Table 1.** Estimates of the parameters in the two-layer model. $\varsigma_1$ is the bulk electron density of the suspensions. $D$ is the diameter of the colloidal particles. $l_1$, $l_2$ are the thicknesses of the colloidal monolayer $\rho_1$, the low-density layer $\rho_2$, respectively. $\varsigma_1$ and $\rho_j$ are normalized to $\varsigma_w =$ 3.33×10$^{29}$ e$^-$/m$^3$. $\sigma_0 = 4.1 \pm 0.2$ Å is the interfacial width between bulk hexane and the low-density layer. $\sigma_1$ is that between the low-density layer and the colloidal monolayer, and $\sigma_2$ is that between bulk of the electrolyte and the colloidal monolayer.

| $D$(Å) | $\varsigma_1$ | $l_1$ (Å) | $l_2$ (Å) | $\sigma_1$ (Å) | $\sigma_2$ (Å) | $\rho_1$ | $\rho_2$ | $\chi^2$ |
|---|---|---|---|---|---|---|---|---|
| 50 | 1.08 | 86 ± 6 | 59 ± 4 | 16 ± 2 | 11 ± 2 | 1.03 ± 0.01 | 1.14 ± 0.01 | 15 |
| 70 | 1.15 | 65 ± 8 | 60 ± 2 | 14 ± 2 | 11 ± 1 | 1.06 ± 0.01 | 1.26 ± 0.01 | 24 |
| 120 | 1.20 | 250 ± 20 | 90 ± 10 | 140 ± 5 | 27 ± 1 | 1.14 ± 0.01 | 1.26 ± 0.01 | 19 |

**Table 2.** Estimates of the parameters in the three-layer model with a compact layer. $D$ is the diameter of the colloidal particles. $l_1$, $l_2$, and $l_3$ are the thicknesses of the compact layer $\rho_1$, the low-density layer $\rho_2$ and the colloidal monolayer $\rho_3$, respectively. $\rho_j$ is normalized to $\varsigma_w = 3.33 \times 10^{29}$ e$^-$/m$^3$. $\sigma_0 = 4.1 \pm 0.2$ is the interfacial width between bulk hexane and the low-density layer. $\sigma_2$ is that between the low-density layer and the colloidal monolayer, and $\sigma_3$ is that between bulk of the electrolyte and the colloidal monolayer.

| $D$(Å) | $l_1$ (Å) | $l_2$ (Å) | $l_3$ (Å) | $\sigma_2$ (Å) | $\sigma_3$ (Å) | $\rho_1$ | $\rho_2$ | $\rho_3$ | $\chi^2$ |
|---|---|---|---|---|---|---|---|---|---|
| 50 | 10 ± 7 | 86 ± 6 | 58 ± 1 | 18 ± 1 | 12 ± 1 | 1.10 ± 0.01 | 1.03 ± 0.01 | 1.14 ± 0.01 | 8.7 |
| 70 | 20 ± 6 | 65 ± 8 | 60 ± 2 | 16 ± 1 | 12 ± 1 | 1.13 ±0.01 | 1.06 ± 0.01 | 1.27 ± 0.01 | 17 |
| 120 | 7 ± 5 | 220 ±6 | 108 ±3 | 144 ±5 | 27 ± 1 | 1.5 ± 0.3 | 1.14 ± 0.01 | 1.26 ± 0.01 | 6.5 |

**Table 3.** Three-layer model of the transition layer at n-hexane/silica sol interface (see Fig. 7). The concentrations of the silica particles in the loose monolayer, $c_3$, and in the low-density layer, $c_2$, are relative to the bulk concentration, $c_b$ (see Ref. 27). $\Gamma$ is the integral number of electrons in the thin layer. $\Gamma^+$ is the surface concentration of sodium ions in the compact layer.

| $D$(Å) | $c_3/c_b$ | $c_2/c_b$ | $\Gamma$ (10$^{20}$ m$^{-2}$) | $\Gamma^+$ (10$^{18}$ m$^{-2}$) |
|---|---|---|---|---|
| 50 | 1.9 | 0.4 | 1.2 – 6 | 2 ± 1 |
| 70 | 1.7 | 0.4 | 5 – 10 | 4 ± 2 |
| 120 | 1.3 | 0.7 | 1.1 – 5 | 4 ± 3 |

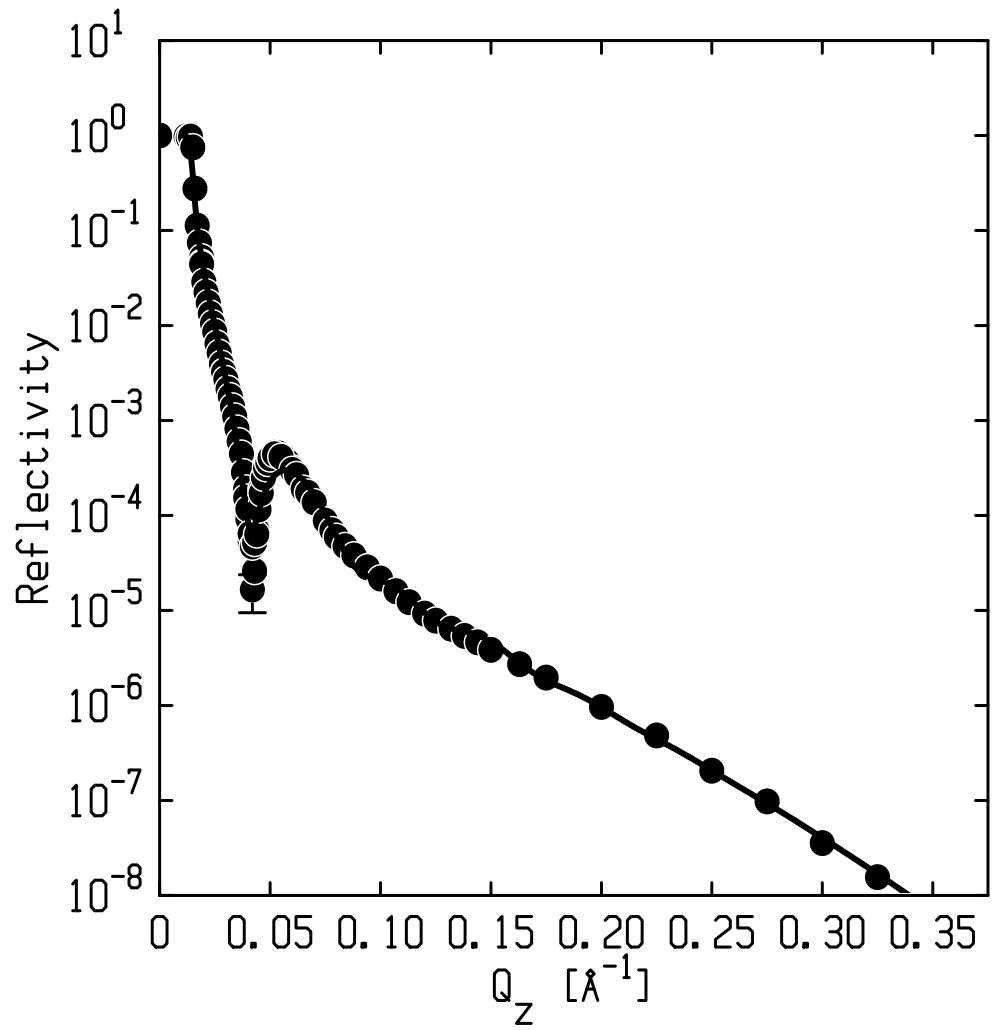

Fig. 1 Tikhonov

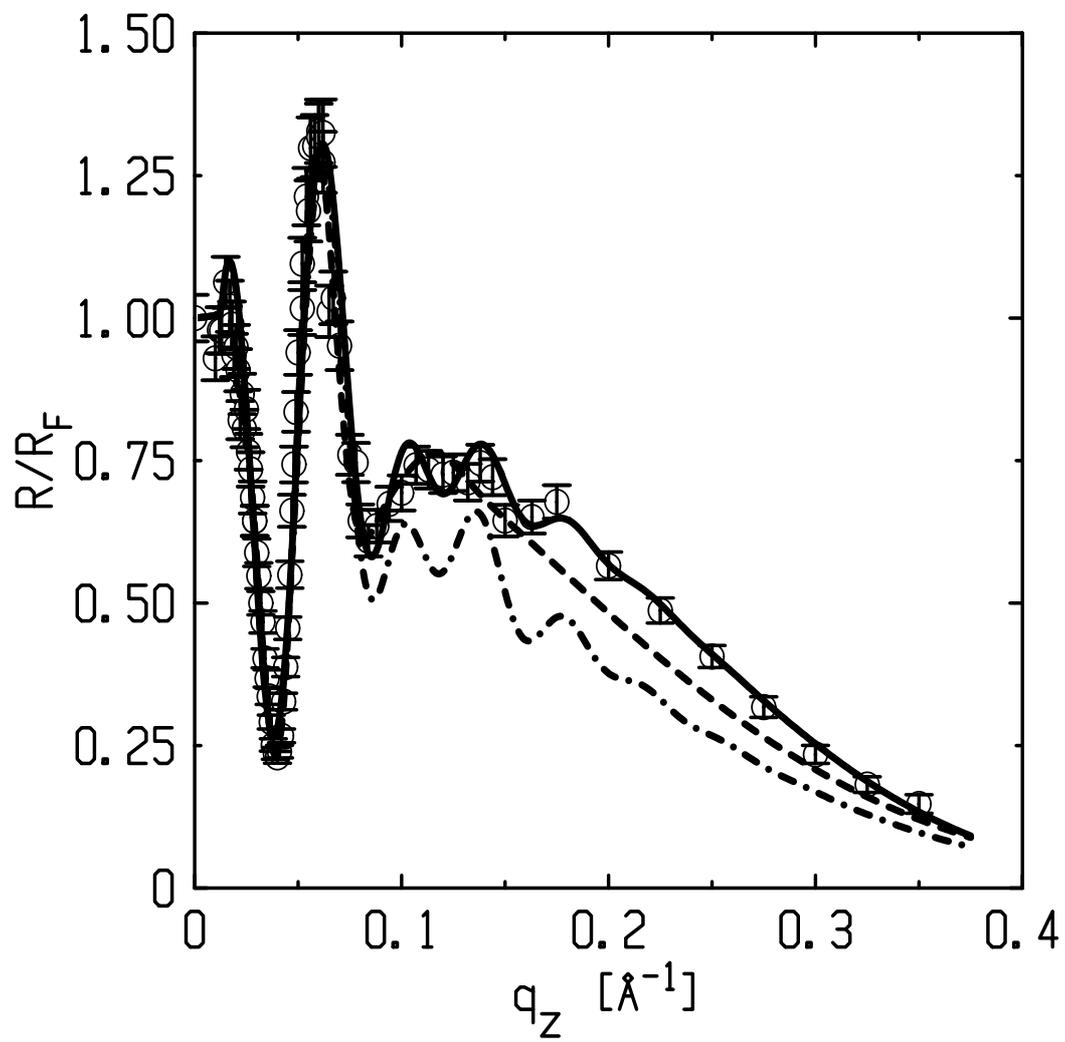

Fig. 2 Tikhonov

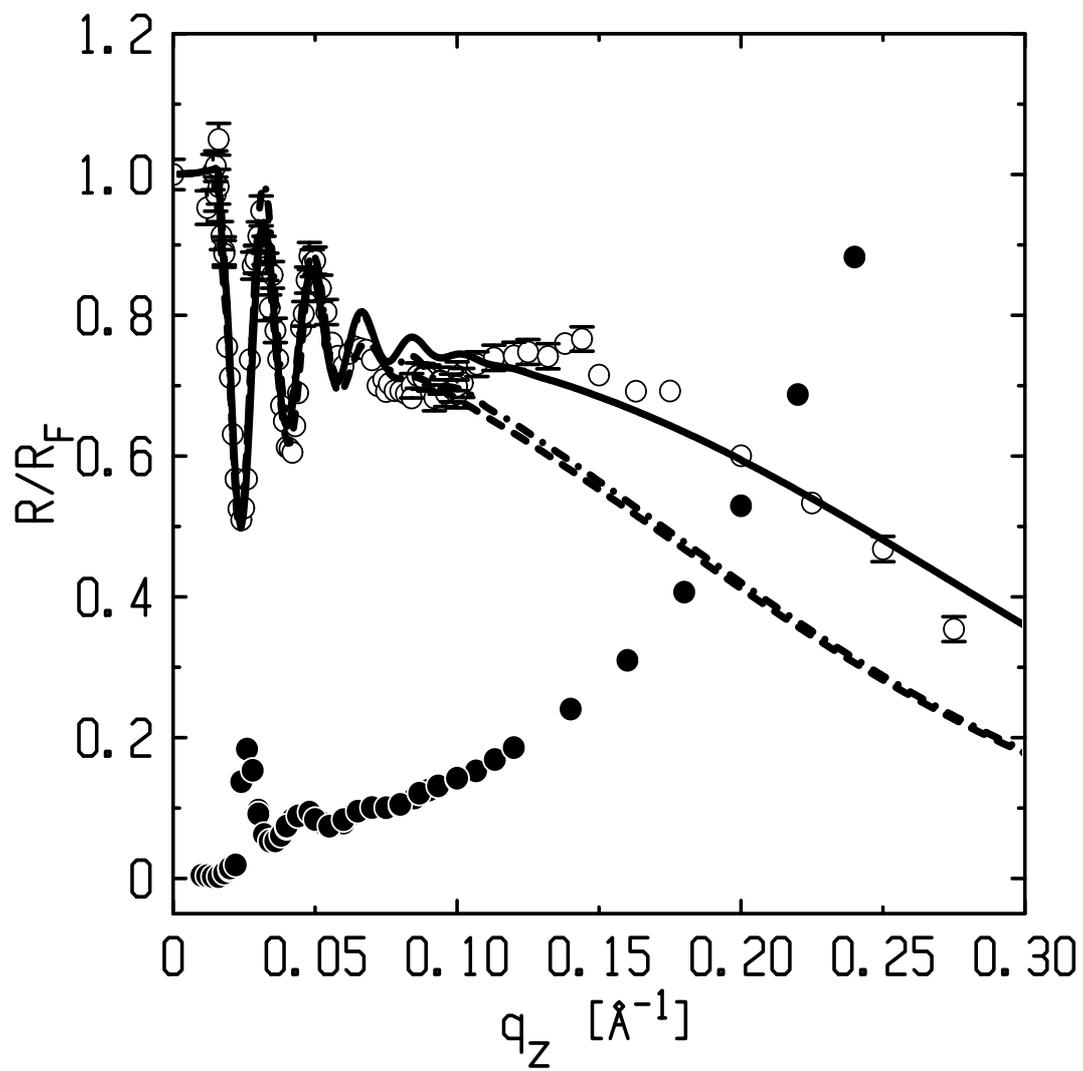

Fig. 3 Tikhonov

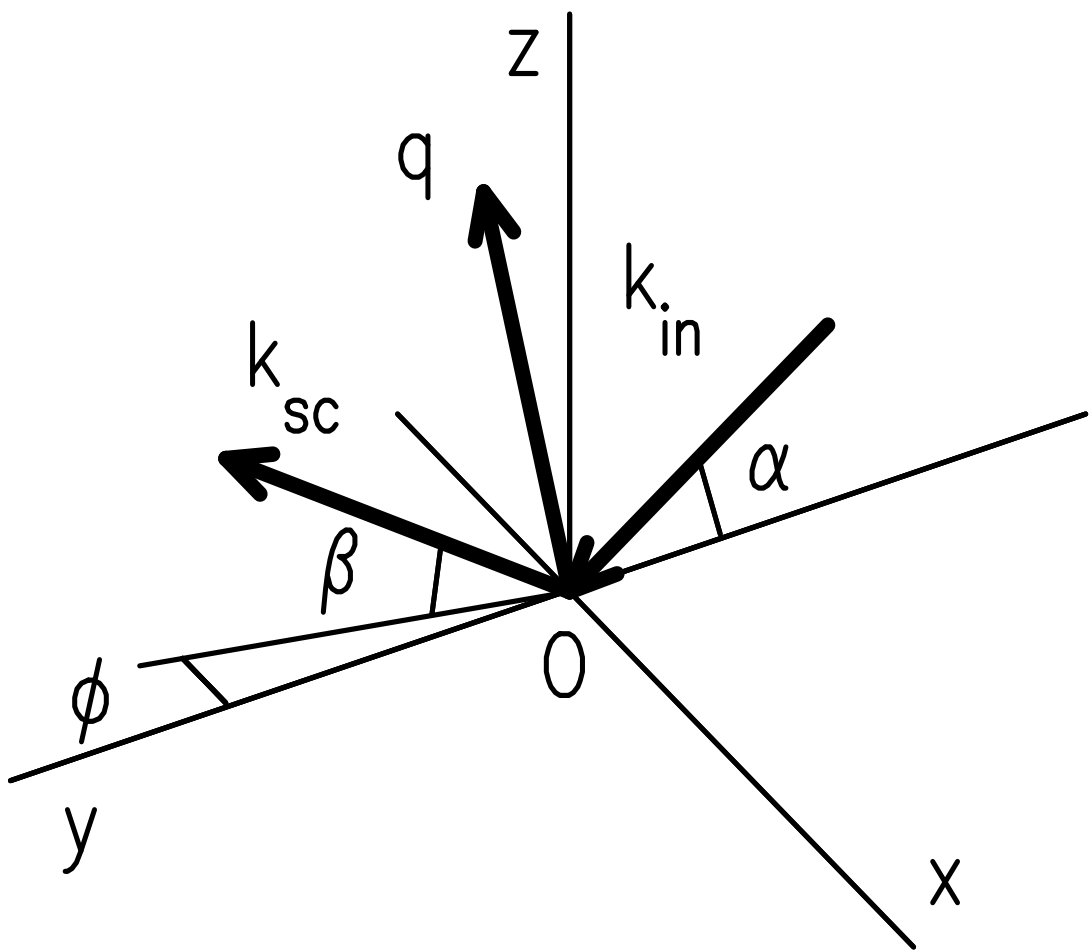

Fig. 4 Tikhonov

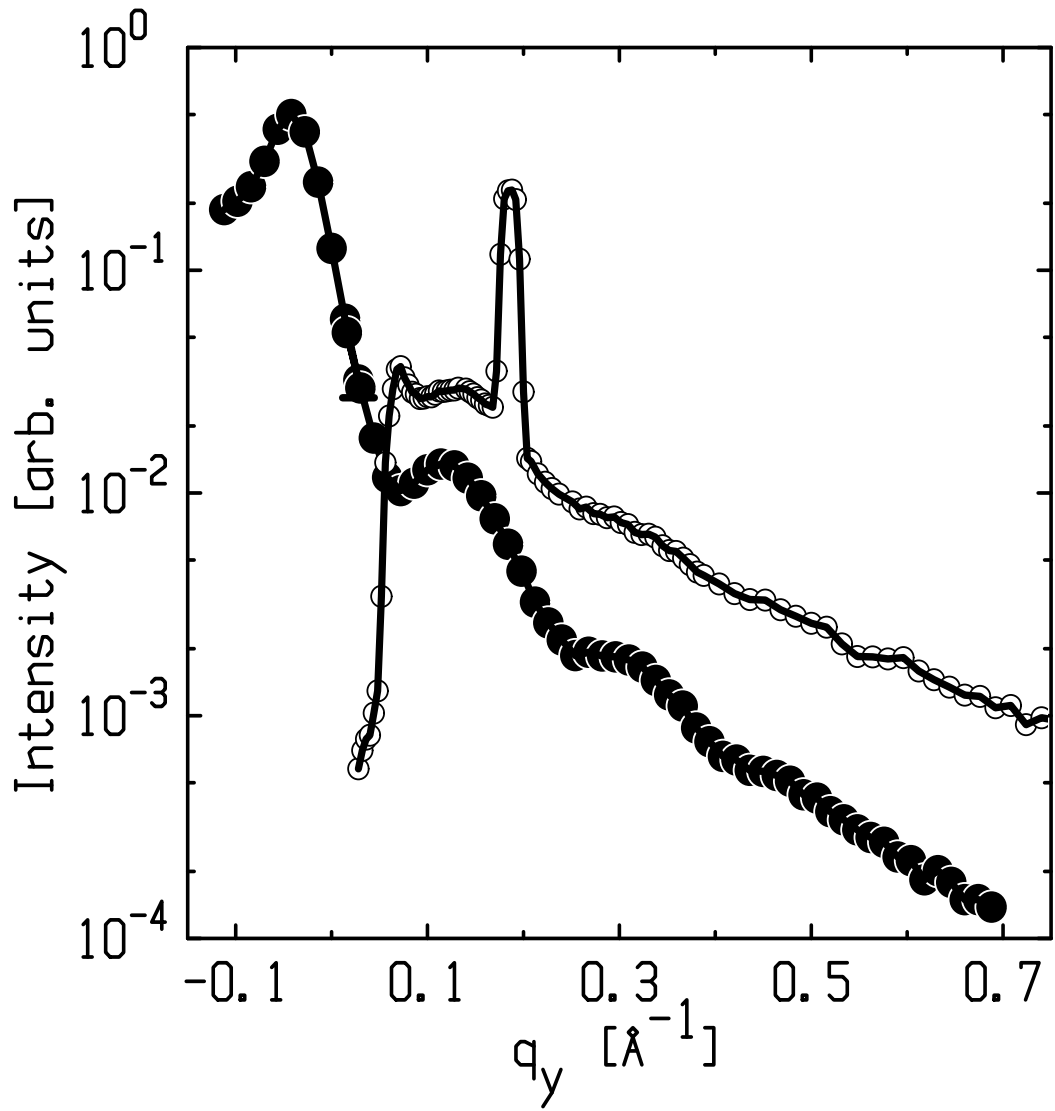

Fig. 5 Tikhonov

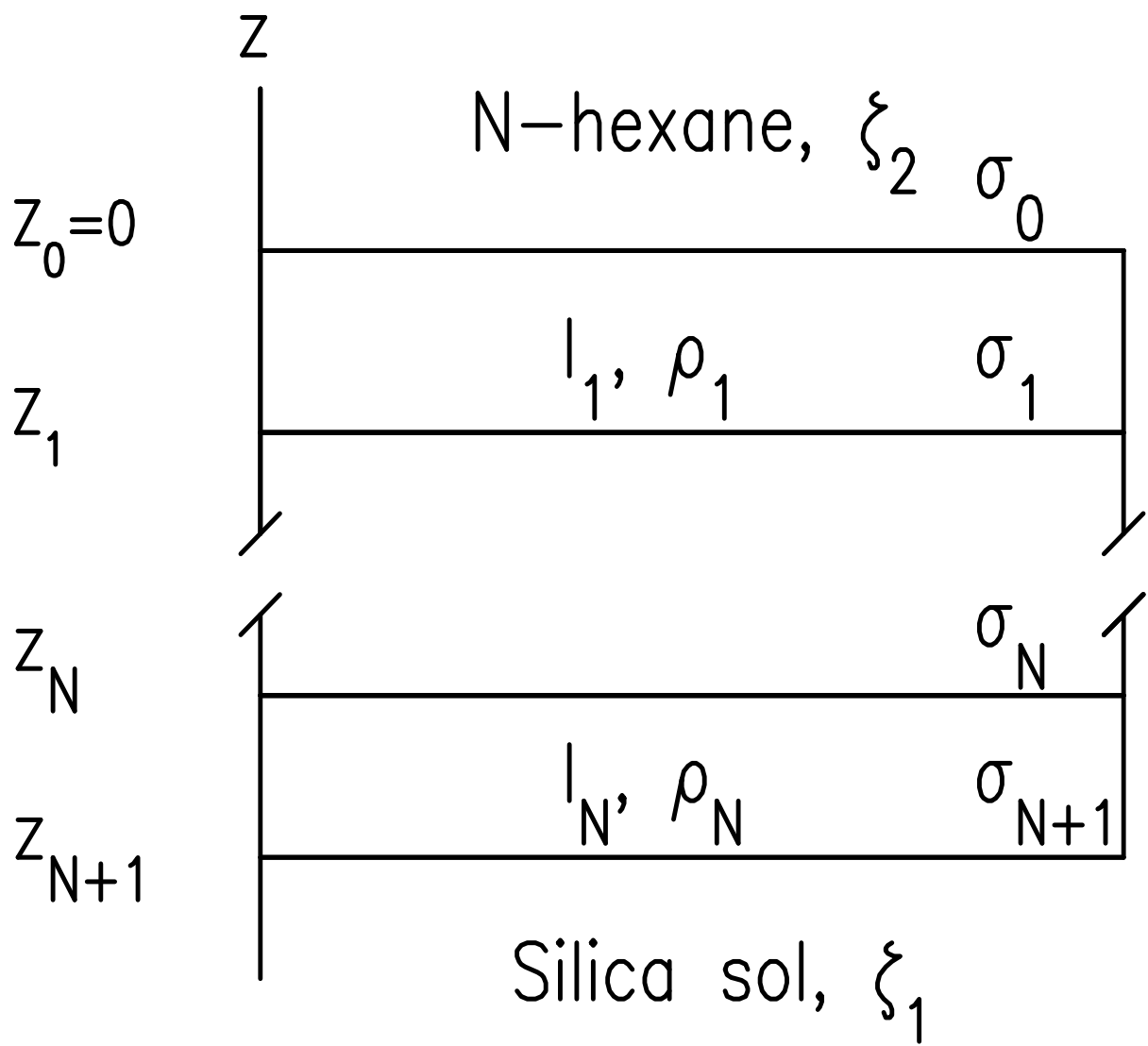

Fig. 6 Tikhonov

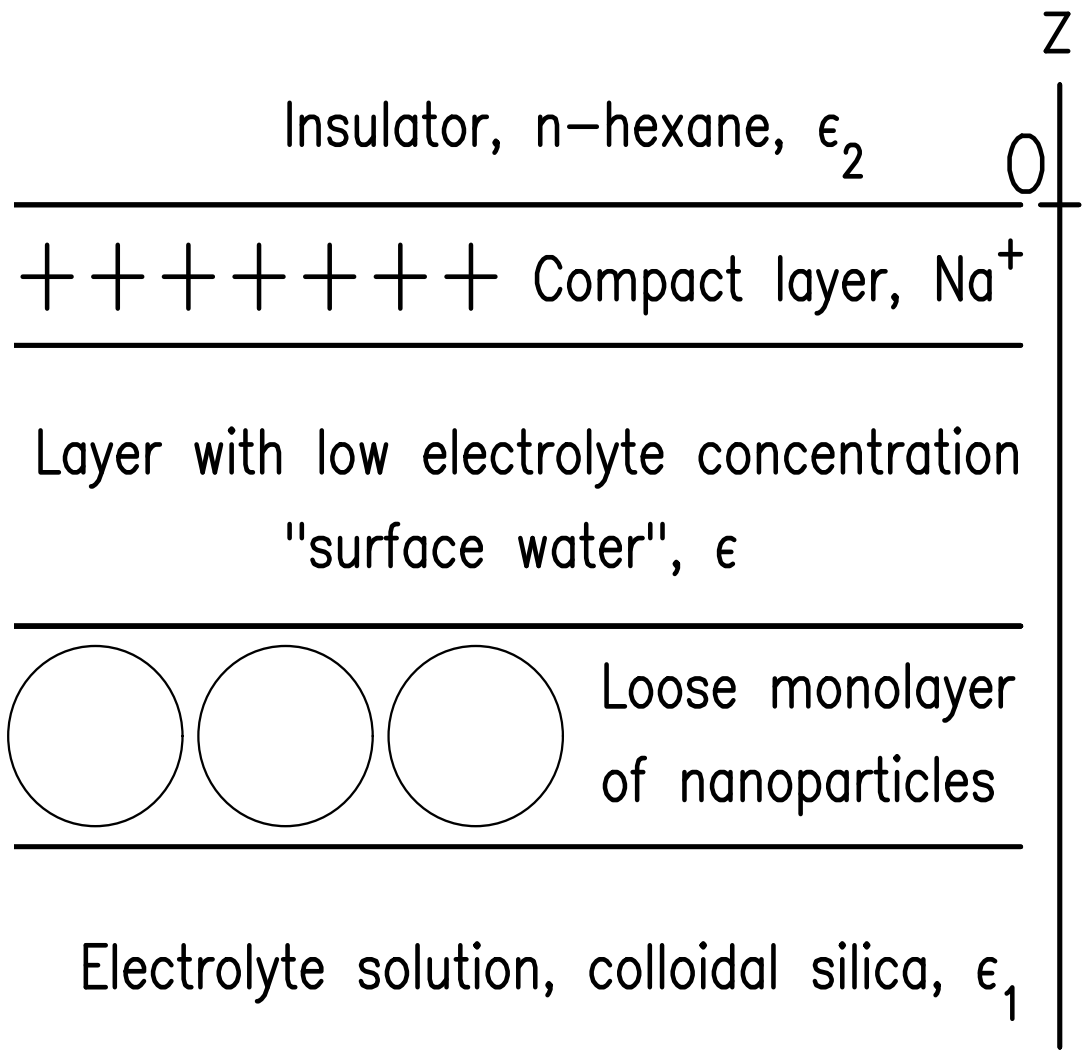

Fig. 7 Tikhonov

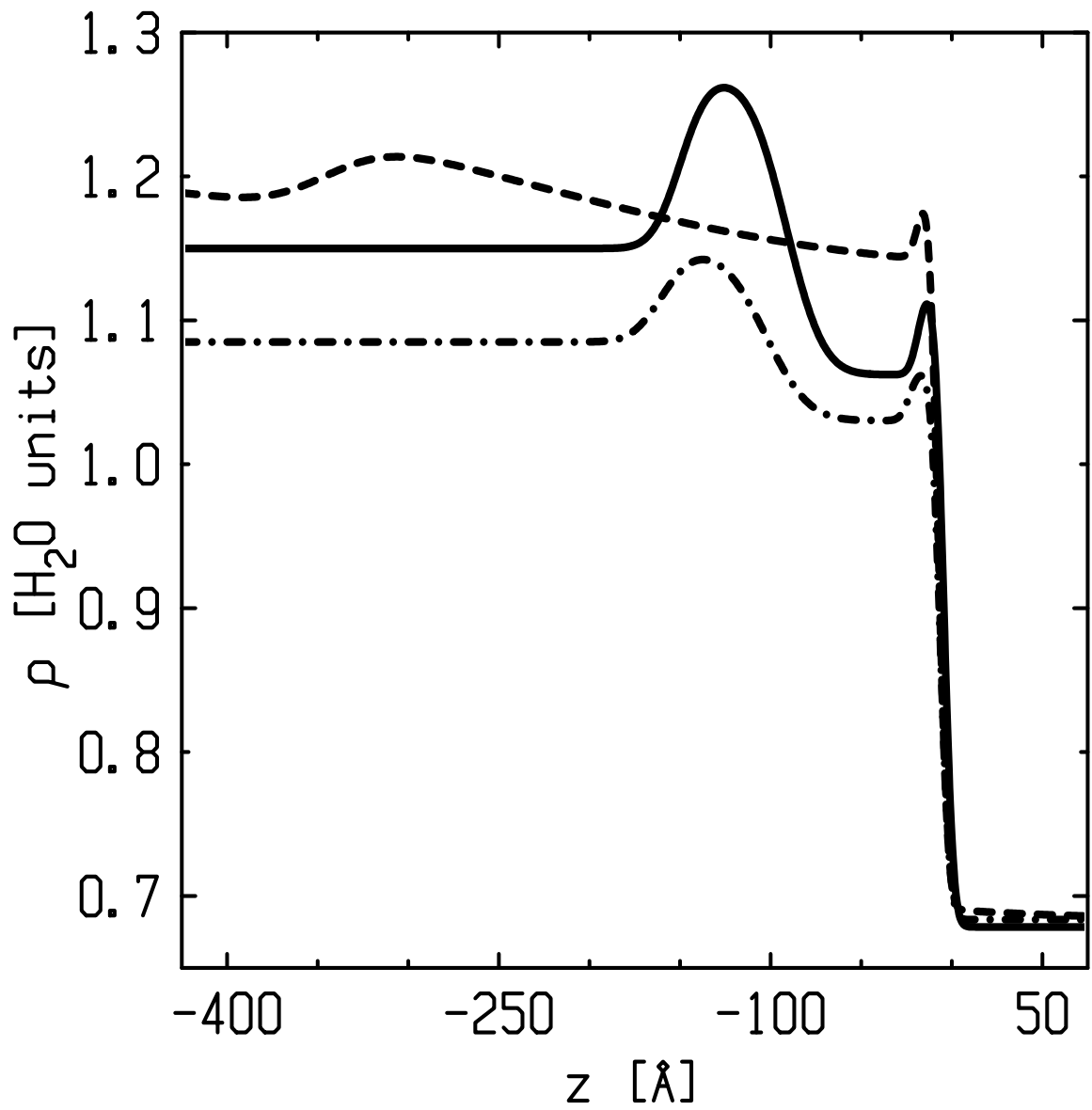

Fig. 8 Tikhonov

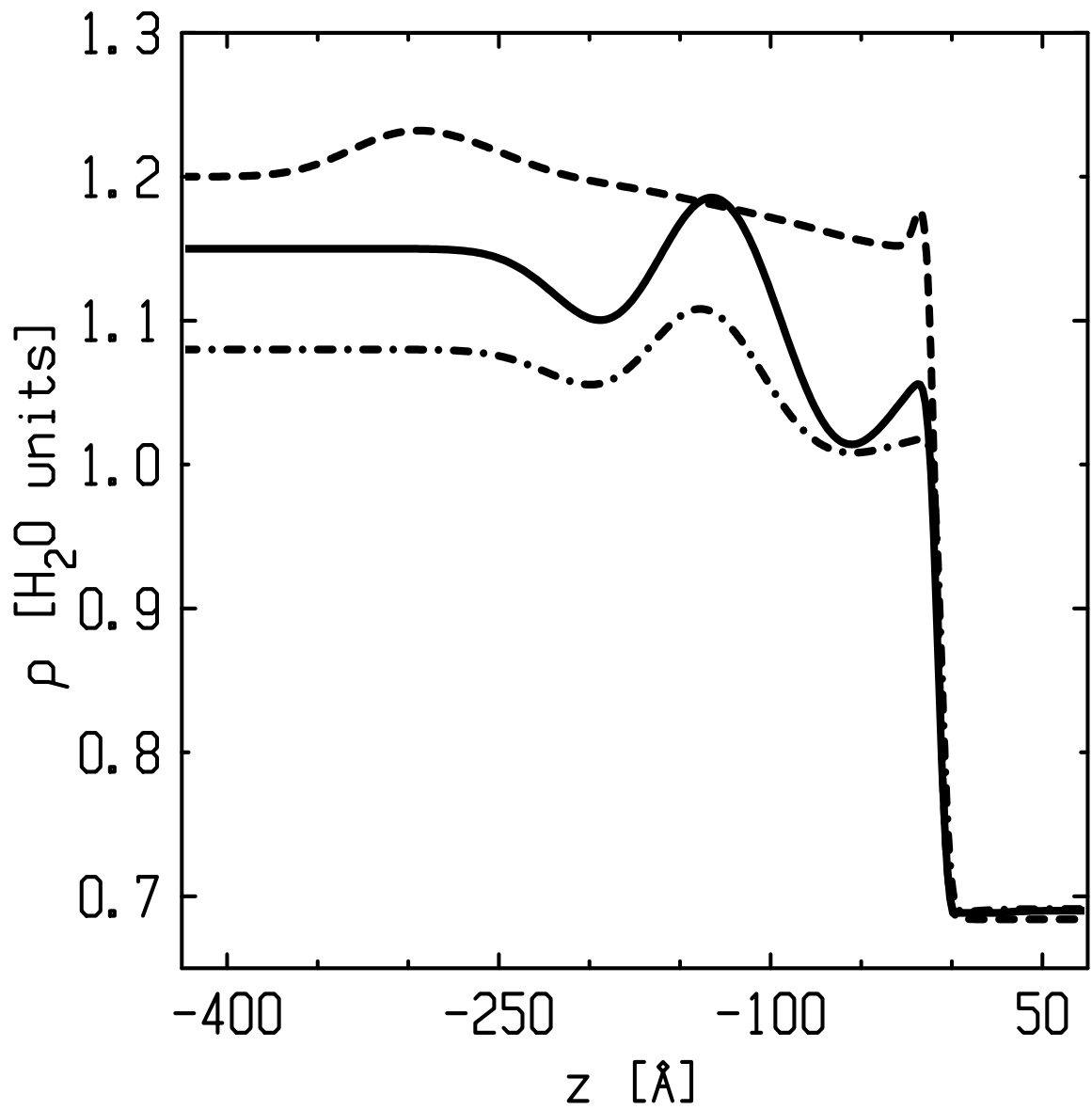

Fig. 9 Tikhonov